\documentstyle[12pt,prl,aps,epsf]{revtex}

\newcommand{\m}[1]{ { $#1$} }

\newcommand{\Em}{ \em  }
\newcommand{\beq}{ \begin{eqnarray}}
\newcommand{\eeq}{\nonumber\end{eqnarray}}
\newcommand{\numeq}{\end{eqnarray}}
\newcommand{\eqn}[1]{~(\ref{#1})}

\def \M{{ \Lambda}}
\def \f{\tilde{f}}

\def \P{\Psi}
\def\tr{\;\;{\rm tr}\;\;}
\def\Tr{\;\;{\rm Tr}\;\;}

\begin{document}

\title{Symmetries of Large \m{N_c}  Matrix Models for Closed Strings}

\author{ C.-W. H. Lee and S. G. Rajeev\\
{\Em Department of Physics and Astronomy,University of Rochester,Rochester,\\} 
{\Em New York 14627}}

\maketitle

\abstract{
We obtain the symmetry algebra of  multi-matrix models in the planar large 
\m{N_c} limit. We use this algebra to associate these matrix models with quantum spin chains.
In particular, certain multi-matrix models are exactly solved by using known results of 
solvable spin chain systems.}
\pagebreak 

Quantum systems whose degrees of freedom are matrices appear in several areas 
of mathematics and physics; for example, Yang-Mills theory
\cite{thwi}, \cite{thorn}, \cite{uym}, \cite{dekl}, 
string theory\cite{douglas}, \cite{bergth} and  \m{M}-theory\cite{feka}, 
\cite{susskind}.
 Of particular interest is the limit  as the dimension \m{N_c} 
of the matrices goes to 
infinity. In this limit the dynamics is expected to simplify; for
example, the quantum fluctuations  of  the invariants are  of order
\m{1/N_c}
.  The algebra of invariant observables becomes a Poisson
algebra discovered in Ref.\cite{tura}. 
For  the general large \m{N_c}
limit,
these Poisson brackets are
very non-linear. The {\Em planar} large \m{N_c} limit is equivalent to
a further approximation  that replaces this  Poisson algebra with a Lie
algebra.
In this paper we will  describe  this Lie algebra of
observables of the matrix model in the planar limit, by a direct
argument.

As an illustration of the power of this new symmetry algebra,
we will use it  to solve some matrix models in the large \m{N_c} limit.
More precisely, we will map certain matrix models to quantum spin
chains and use results from the theory of spin chains to solve them.
 This is reminiscent of the work \cite{douglas} that connects some
integrals  over
finite chains  of  matrices   with {\Em classical} integrable systems.
 From this  point of view,  our result is that certain {\it path
integrals}  over 
matrices  can be mapped into {\Em quantum} integrable systems. However
we will mostly use the  canonical formulation rather than the path
integral formulation of these systems.

We will study a class of matrix models whose degrees of freedom are a set of 
matrix-valued bosonic  variables \m{a^{\mu}_\nu(i),
a^{\dag\mu}_\nu(i)} satisfying the canonical commutation relations 
\m{[a^\mu_\nu(i),a^\rho_\sigma(j)]=[a^{\dag\mu}_\nu(i),a^{\dag\rho}_\sigma(j)]=0} and 
\m{[a^\mu_\nu(i),a^{\dag\rho}_\sigma(j)]=
	\delta(i,j)\delta^\mu_\sigma\delta^\rho_\nu .}
 Here, \m{\mu,\nu=1,2,\ldots} or \m{N_c}. The positions of the indices indicate the transformation properties under \m{U(N_c)}:
\m{
	a^\mu_\nu\rightarrow g^\mu_\rho g^{*\sigma}_\nu a^\rho_\sigma
}
etc.  The degree of freedom labelled by the indices \m{\mu,\nu} etc. will 
be called `color' in analogy with quantum chromodynamics (QCD). Indeed our matrix 
model can be thought of as a regularized version of pure  QCD, with the  
variables \m{a,a^{\dag}} representing gluons.
The indices  \m{i=1,\cdots, \M} describe the 
 degrees of freedom (other than color) of the system. The Hamiltonian
(along with all other observables) will be required to be color
invariant, i.e., invariant under the adjoint action of  \m{U(N_c)} on \m{a} and
 \m{a^{\dag}}. 

The path
integral over  matrix-valued functions of time,
\m{P^\mu_\nu(j,t), Q^\mu_\nu(j,t)} with Lagrangian
\beq
	L(P,Q)= \sum_{j=1}^{\M}P^\mu_\nu(j,t){d\over dt} Q^{\nu}_\mu(j,t)-H(P(t),Q(t))
\eeq
gives an equivalent theory, with the identifications
\m{a^\mu_\nu(j)=Q^\mu_\nu(j)+iP^\mu_\nu(j),a^{\mu\dag}_\nu(j)=
Q^\mu_\nu(j)-iP^\mu_\nu(j)}; but 
   the canonical  formulation is more convenient for our purposes.

 Define the vacuum state of the representation of these relations 
by 
\m{
	a|0>=0
}.
In the limit of large \m{N_c} the color invariant states of the
 system are the `closed string' (or `glueball') states   such as 
\beq
 \Psi^{(K)} = N_c^{-c/2} a^{\dagger\mu_1}_{ \mu_2} (k_1) 
   a^{\dagger\mu_2}_{ \mu_3} (k_2) \cdots a^{\dagger\mu_c}_{ \mu_1} (k_c)     
    |0\rangle .
\label{1.0.2}
\eeq
Here strings of indices  are denoted by
capital letters. For example,  \m{K} stands for  \m{k_1,\cdots, k_c}. 
The state is invariant under cyclic
permutations; the equivalence class of permutations related to \m{K} by cyclic
permutations is denoted by  \m{(K)}.

The   operators that dominate the large \m{N_c} limit  are 
\beq
    g^I_{J} & \equiv & N_c^{-(a+b-2)/2} a^{\dagger\mu_1}_{ \mu_2} (i_1)
   a^{\dagger\mu_2}_{ \mu_3} (i_2) \cdots a^{\dagger\mu_a}_{ \nu_b} (i_a) 
   \nonumber \\
   & & a_{\nu_{b-1}}^{ \nu_b} (j_b) a_{\nu_{b-2}}^{ \nu_{b-1}} (j_{b-1}) \cdots
   a_{\mu_1}^{ \nu_1} (j_1).
\eeq
(Notice the reversal of order in the indices in the string \m{J}; this  
definition serves to simplify some later equations.)
All observables of a matrix model which survive in the large \m{N_c} limit--
the Hamiltonian of regularized QCD for example -- are  linear combinations of
 such operators. These states and operators were previously studied
in Ref. \cite{thorn}, where an  elegant application  to large  \m{N_c} QCD
is described .

The  factors of \m{N_c}  have been chosen to obtain the `planar' limit;
it is so called because in perturbation theory,
 the Feynman diagrams that survive can be drawn on a plane. There are
other ways of taking the large \m{N_c} limit, but the planar limit is
the simplest.

In the limit as \m{N_c\to\infty} these operators will map single closed string states
to linear combinations of single closed string states (``glueballs'') : 
\beq
   g^I_{J} \P^{(K)} = \delta^K_{(J)} \P^{(I)} + \sum_{K_1 K_2 = (K)} 
   \delta^{K_1}_J \P^{(I K_2)}. 
\label{5.1}      
\eeq
This is the key simplification of the planar limit.
(To  higher orders in the 
\m{1/N_c} expansion, there will be terms that correspond to splitting  a 
glueball into several glueballs.)
Here, \m{\delta^K_{(J)}} is equal to the number of different cyclic permutations of $J$ 
such that each permuted sequence is identical to $K$. Also, in the second term we sum over all ways of 
splitting the sequence \m{(K)} into {\it non-empty} 
 subsequences \m{K_1} and \m{K_2}.
A graphical representation of \eqn{5.1} is given in Fig. 1.

The operators \m{g^I_J} are  like matrices except that they  operate
 on the
 space of cyclically symmetric tensors. We will call them `cyclix' operators.
The product \m{g^I_J g^K_L} of two of the above operators is not a
 finite linear
 combination of  the \m{g}'s themselves. But the commutator is indeed such a
 finite linear combination: {\it finite linear combinations of the operators \m{g^I_J} form  a Lie algebra}. 
 (By finite linear cobinations we mean a sum over all  sequences of 
indices \m{I} and \m{J}, of the form 
\m{\sum c^J_Ig^I_J}, such that only a finite number of the coefficients   \m{c^J_I}
are non-zero.) The discovery of this Lie algebra is our main result. We
 will see that it has powerful consequences: for example we can solve
 some matrix models exactly using this newly discovered dynamical symmetry.

Before we describe the commutation relations between two \m{g^I_J}'s ,  it is
convenient to introduce another kind of  operator \m{\f^{(I)}_{(J)}} on closed string states.
The defining equation for these operators is
\m{
	\f^{(I)}_{(J)}\P^{(K)}=\delta^K_{(J)}\P^{(I)}.
}
These are thus the Weyl matrices in the basis \m{\P^{(I)}} of closed string 
states up to constant multiples. Rather than being independent operators, they are 
 in fact just linear combinations of  
\m{g^{I}_{J}}:
\m{
   \f^{(I)}_{(J)}  =  g^I_J - \sum_{k=1}^{\M} g^{I k}_{J k}
}
and 
\m{ 
   \f^{(I)}_{(J)} =  g^I_J - \sum_{k=1}^{\M} g^{k I}_{k J}. 
}
The two different ways of writing  \m{\f^{(I)}_{(J)}} imply 
that the operators \m{g^I_J} are not linearly independent. 

Now we can state the commutation relations of our Lie algebra:
\beq
   \lefteqn{\left[ g^I_{{J}}, g^K_{{L}} \right] =
   \delta^K_J g^I_{{L}} + \sum_{J_1 J_2 = J} \delta^K_{J_2} 
   g^I_{{J_1 L}} + \sum_{K_1 K_2 = K} \delta^{K_1}_J
   g^{I K_2}_{{L}} } \nonumber \\
   & & + \sum_{\begin{array}{l}
		  J_1 J_2 = J \\
		  K_1 K_2 = K
	       \end{array}}
   \delta^{K_1}_{J_2} g^{I K_2}_{{J_1 L}} 
   + \sum_{J_1 J_2 = J} \delta^K_{J_1} g^I_{{L J_2}} 
   + \sum_{K_1 K_2 = J} \delta^{K_2}_J g^{K_1 I}_{{L}} \nonumber \\
   & & + \sum_{\begin{array}{l}
	       J_1 J_2 = J \\
	       K_1 K_2 = K
	    \end{array}}
   \delta^{K_2}_{J_1} g^{K_1 I}_{{L J_2}} 
   + \sum_{J_1 J_2 J_3 = J} \delta^K_{J_2} g^I_{{J_1 L J_3}} \nonumber \\ 
   & & + \sum_{K_1 K_2 K_3 = K} \delta^{K_2}_J g^{K_1 I K_3}_{{L}} 
   + \sum_{\begin{array}{l}
   		  J_1 J_2 = J \\
   		  K_1 K_2 = K
   	       \end{array}}
   \delta^{K_1}_{J_2} \delta^{K_2}_{J_1} \f^{(I)}_{({L})} \nonumber \\
   & & + \sum_{\begin{array}{l}
   	      J_1 J_2 J_3 = J \\
   	      K_1 K_2 = K
   	   \end{array}}
   \delta^{K_1}_{J_3} \delta^{K_2}_{J_1} \f^{(I)}_{({J_2 L})}
   + \sum_{\begin{array}{l}
   		  J_1 J_2 = J \\
   		  K_1 K_2 K_3 = K
   	       \end{array}}
   \delta^{K_1}_{J_2} \delta^{K_3}_{J_1} \f^{(I K_2)}_{({L})} \nonumber \\
   & & + \sum_{\begin{array}{l}
		  J_1 J_2 J_3 = J \\
		  K_1 K_2 K_3 = K
	       \end{array}}
   \delta^{K_1}_{J_3} \delta^{K_3}_{J_1} \f^{(I K_2)}_{({J_2 L})} 
   - (I \leftrightarrow K, J \leftrightarrow L). 
\eeq
Although it appears complicated when written this way, these commutation
 relations have a rather natural graphical interpretation which we will 
describe  in a longer paper\cite{clstal}. We will call the Lie algebra defined by these 
commutation relations  the `cyclix Lie algebra' or \m{\hat {\underline{C}}_M}.

The above defined \m{\f^{(I)}_{(J)}} span an ideal of this algebra isomorphic 
to the inductive limit of linear algberas, \m{gl_{\infty}}.
(\m{gl_{\infty}} can also be defined as the Lie algebra of matrices
with only 
a finite number of non-zero entries.)

 We can quotient \m{\hat{\underline{C}}_M} 
by this ideal to get another Lie algebra \m{{\underline{C}}_M}, which is the essentially new object
 we have discovered. However  it is only the  extension \m{\hat {\underline{C}}_M}
that has a representation on the space of closed string states. 

In the simplest special case  of a matrix model with just one degree 
of freedom (\m{M=1}),  the algebra \m{{\underline{C}}_1} is just the algebra of (polynomial)
 vector fields  on the circle. \m{\hat{\underline{C}}_1} is then the
extension of 
this algebra by the algebra of finite-rank matrices\cite{Kac}. Perhaps 
then \m{{\underline{C}}_M} can be realized as the Lie algebra of
vector fields
 on a non-commutative manifold.

 We will now show how some large \m{N_c} matrix models can be solved by using this
new symmetry algebra.
Suppose the Hamiltonian of a matrix model is a linear combination 
\m{
	H=\sum_{IJ} h^J_Ig^I_J
}
where \m{h^J_I=0} unless \m{I} and \m{J} have the same number of indices.
( This means that the `gluon number' is a conserved quantity: regularized 
QCD is not of this type.)  Such linear combinations form a subalgebra; let us
 call it \m{\hat{\underline{C}}_M^0}.

{\Em  There is an isomorphism between multi-matrix models whose hamiltonians 
are in \m{\hat{\underline{C}}_M^0} and  quantum spin chains}. Now, there are some well-known
 examples of exactly solved quantum  spin chains; they yield exactly
 solved matrix models.

 More explicitly, consider a  spin chain with \m{\nu} sites: at any site 
\m{a=1},\ldots, or \m{\nu}, there is a variable \m{i_a} (called `spin' for historical 
reasons) that can take the value 1, 2, \m{\ldots}, or \m{\M}. We will impose the periodic 
boundary condition. A basis of states is given by
\m{
	|k_1\cdots k_\nu>.
}

Define the  operator
\beq
	X^i_j(a)|k_1\cdots k_\nu>=\delta^{k_a}_j|k_1\cdots k_{a-1}ik_{a+1}\cdots k_\nu>.
\eeq
This is just the Weyl matrix at site \m{a}.  Let $X^i_j(a) = X^i_j(a - \nu)$ if $a > \nu$.
Now we can check that if \m{I} and \m{J} have the same length \m{b\leq \nu}, 
\beq
	r_\nu(g^I_J) \equiv \sum_{a=1}^{\nu} X^{i_1}_{j_1}(a)X^{i_2}_{j_2}(a+1)\cdots X^{i_b}_{j_b}(a+b-1)
\eeq
satisfies the commutation relations of the algebra \m{\hat{\underline{C}}^0_M}.
If we also set  \m{r_\nu(g^I_J)=0} for \m{b> \nu}, we will have a representation 
\m{r_\nu} of \m{\hat{\underline{C}}^0_M}. The states of the periodic spin chain with
 zero total momentum correspond to  cyclically symmetric tensors which are the states of the matrix model.

With each matrix model whose hamiltonian 
\m{H=\sum_{IJ}h^J_Ig^I_J} is in \m{\hat{\underline{C}}^0_M}, we can associate a quantum spin chain with the 
Hamiltonian 
\beq
	H^{\rm spin} =\sum_{IJ} h^J_I \sum_{a=1}^{\nu} X^{i_1}_{j_1}(a)X^{i_2}_{j_2}(a+1)\cdots X^{i_b}_{j_b}(a+b-1).
\eeq
Thus matrix models conserving the gluon number correspond to 
 quantum spin systems with interactions involving 
neighborhoods of spins \m{\{a, a+1, a+2,\ldots, a+b-1\}}.

Let us look at some examples of solvable spin models and their associated matrix models. 
The simplest solvable quantum spin chain is perhaps the Ising model 
\cite{onsager},\cite{kogut}:
\beq
	H^{\rm spin}_{\rm Ising}=\sum_{a=1}^{\nu} \tau^z(a) + \lambda \sum_{a=1}^{\nu} \tau^x(a) \tau^x(a+1).
\eeq
Here \m{\tau^{x,y,z}_a} are Pauli matrices at site \m{a}. Let the states 1 and 2 in the matrix model correspond to
the spin-up and spin-down states in the Ising model.  Using the fact that 
\m{
	\tau^z_a = X^1_1(a) - X^2_2(a)
}
and 
\m{
 \tau^x_a + i \tau^y_a = 2X^1_2(j),
}
we get the corresponding element in \m{\hat{\underline{C}}_2^0}:
\beq
	H^{\rm matrix}_{\rm Ising}=g^1_1-g^2_2 + \lambda[g^{22}_{11}+g^{21}_{12}+g^{12}_{21}+g^{11}_{22}].
\eeq
This is the large \m{N_c} limit of the matrix model with the Hamiltonian 
\beq
	H^{\rm matrix}_{\rm Ising}&=&\tr [a^{\dag}(1)a(1)-a^{\dag}(2)a(2)]
+{\lambda \over N_c}\tr [a^{\dag }(2)a^{\dag }(2)a(1)a(1)\cr
& & +
a^{\dag }(2)a^{\dag }(1)a(2)a(1)+a^{\dag }(1)a^{\dag }(2)a(1)a(2)+a^{\dag }(1)a^{\dag }(1)a(2)a(2)].
\eeq

Our results, along with known results of the Ising spin chain \cite{kogut} give the
 spectrum of this matrix model in the large \m{N_c} limit:
\beq
	E(n_p,\nu)=-2\sum_{p=-\nu}^\nu 
		\Bigg(1+2\lambda \cos\big[{2\pi p\over 2\nu +1} \big] +
					\lambda^2\Bigg)^{1/2}n_p
\eeq
where \m{\nu} is any positive integer and \m{n_p = 0} or 1.
Also, we must impose the condition  \m{\sum_{p=-\nu}^\nu n_p p=0} to
get cyclically symmetric states.
In particular we see that the value \m{\lambda=1} is the critical value of the
 matrix model at which the spectum (in the planar limit) is that of a
massless  free fermion field on a  lattice.

It is interesting to ask whether  the symmetries of the Ising spin chain
can be understood within our formalism . Recall that 
\cite{onsager},\cite{dogrda} the solvability of the Ising model is due
to the existence of an infinite number of conserved quantities. They
form an infinite-dimensional Lie algebra,
the Onsager algebra. This is the Lie algebra generated by iterating
commutators of two operators \m{H_0} and \m{V} satisfying
\m{
	[H_0,[H_0,[H_0,V]]]=16[H_0,V]
}
and 
\m{
[V,[V,[V,H_0]]]=16 [V,H_0].
}
For the Ising model,
\m{
	H_0=H=g^1_1-g^2_2}
and  
\m{
 V=g^{22}_{11}+g^{21}_{12}+g^{12}_{21}+g^{11}_{22}.
}
Clearly, the Onsager algebra is  a subalgebra of \m{{\underline{C}}_M^0}. In
particular, all conserved quantities of the Ising model are contained in our 
cyclix Lie algebra. It is not known whether this Ising  matrix model is solvable  for
 an arbitrary  finite value of \m{N_c}.

To every solved spin chain there is thus a corresponding solved matrix model. 
Instead of a comprehensive list, we  are just going to give a few illustrative
examples.

The generalization of the Ising model with the Hamiltonian \cite{honecker}
\beq
	H^{\rm spin}_{\rm GI}=\sum_{a=1}^{\nu} \tau^z_a+\lambda\sum_{a=1}^{\nu} 
	[\tau^x_a\tau^x_{a+1}+v\{\tau^x_a \tau^y_{a+1}- \tau^y_a \tau^x_{a+1}\}]
\eeq 
also has the Onsager algebra as a dynamical symmetry. It corresponds
to the element 
\beq
	H^{\rm matrix}_{\rm GI}&=&g^1_1-g^2_2+
\lambda[g^{22}_{11}+(1-2iv)g^{21}_{12}+(1+2iv)g^{12}_{21}+g^{11}_{22}]
\eeq
of the cyclix Lie algebra and hence to the exactly solvable matrix model
\beq
	H^{{\rm matrix}}_{{\rm GI}}&=&
\tr [a^{\dag}(1)a(1)-a^{\dag}(2)a(2)]
 +{\lambda\over N_c}\tr [a^{\dag }(2)a^{\dag }(2)a(1)a(1)\cr
 & & +
(1-2iv)a^{\dag }(2)a^{\dag }(1)a(2)a(1)+
(1+2iv)a^{\dag }(1)a^{\dag }(2)a(1)a(2)\cr
& & +a^{\dag }(1)a^{\dag }(1)a(2)a(2)].
\eeq

The XYZ model\cite{XYZ} with the Hamiltonian  
\beq
	H^{{\rm spin}}_{{\rm XYZ}}=\sum_{a=1}^{\nu} \tau^z_a\tau^z_{a+1}-
\lambda\sum_{a=1}^{\nu}[\tau_a^x\tau^x_{a+1}+v\tau^y_a\tau^y_{a+1}]
\eeq
is a generalization of the Ising model in another direction.
The corresponding element in the cyclix algebra is 
\beq
	H^{{\rm matrix}}_{{\rm XYZ}}=g^{11}_{11}-g^{12}_{12}-g^{21}_{21}+g^{22}_{22}-
\lambda[(1-v)(g^{22}_{11}+g^{11}_{22})+(1+v)(g^{21}_{12}+g^{12}_{21})].
\eeq
A special case of this, the equivalence of a matrix model to the XXZ
model, was found in \cite{klebanovsusskind}.

The above correspondence between spin chains and matrix models is not restricted
to the case \m{{\cal M}=2}.  The chiral Potts model \cite{chPotts} has the
Hamiltonian
\beq
	H^{{\rm spin}}_{{\rm CP}}=\sum_{a=1}^{\nu} \sum_{k=1}^{\M -1}[\tilde \alpha_k Q_a^k+\lambda 
\alpha_k P_a^kP^{\M-k}_{a+1}]
\eeq
where \m{\alpha_k,\tilde\alpha_k} are constants. Also, \m{P_a} and
\m{Q_a} are 
generalized  spin matrices at site \m{a}: 
\m{
	Q=\;{\rm diag}\;\;\big(1,\omega,\omega^2,\cdots
\omega^{\M-1}\big)
}
and \m{P} is defined by \m{PQ=\omega QP}. Here, \m{\omega=e^{2\pi i/  \M}}. 
This model is exactly solvable and corresponds to the element 
\beq
   H^{{\rm matrix}}_{{\rm CP}}=\sum_{k=1}^{\M-1}[\tilde\alpha_k\sum_{j=1}^{\nu}\omega^{k(j-1)}g^j_j+
   \lambda\alpha_k\sum_{j_1, j_2=1}^{\nu} g^{j_1 + k, \; j_2 - k}_{j_1 j_2}]
\eeq
where $j_1 + k$ should be replaced with $j_1 + k - \Lambda$ if $j_1 + k > \Lambda$ and $j_2 - k$ should be replaced
with $j_2 + \Lambda - k$ if $j_2 - k \leq 0$ in $g^{j_1 + k, j_2 - k}_{j_1 j_2}$ of the cyclix algebra.

The problem of finding the partition function of the spectrum of a
hamiltonian  \m{H} is equivalent
to evaluating the path integral over paths of period \m{T}:
\m{
	\Tr e^{-iHT}=\int D[P]D[Q]e^{i\int_0^T[\sum_j\tr P(j)\dot
Q(j)-H(P(j),Q(j))]dt}.
}
where \m{H(P,Q)} is obtained by substituting \m{a=Q+iP, a^{\dag}=Q-iP}
into \m{H} as  described previously. By applying this transcription to
the above systems, we can obtain  path integrals over
matrices  which can be evaluated exactly in the  planar large \m{N_c}
limit. We wont give explicit expressions   to keep the
paper short.

In addition to integrable matrix models associated with quantum spin chain models, we have also formulated
models for QCD in terms of elements of the cyclix algebra \cite{clstal}.
We have also found the analog of the cyclix algebra suitable for studying open strings 
(`meson states') \cite{opstal}; the  supersymmetric extension has also been
constructed \cite{sustal}.  The former is of interest in spin chains with open boundary 
conditions and QCD with quarks, and the latter in \m{M}-theory.

\pagebreak
\begin{figure}
\epsfxsize=4in
\centerline{\epsfbox{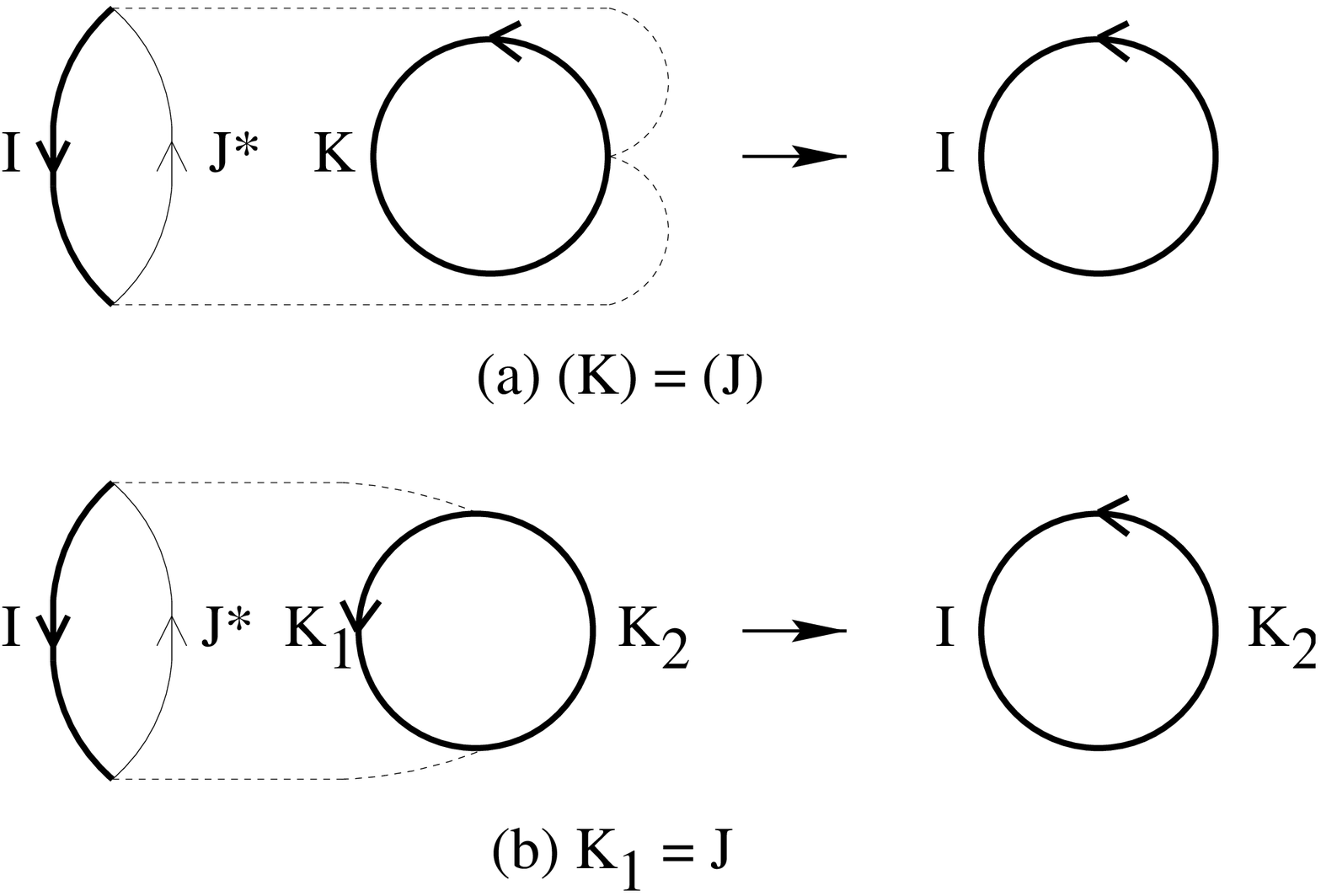}}
\caption{{\em The action of a gluonic operator on a single glueball state.}
The gluonic operator \m{g^I_J} searches for a substring of \m{K} that
agrees with \m{J}. If found, it replaces each such substring 
with \m{I}; otherwise, we get  zero. Here, $J^*$ denotes the reverse of the
sequence \m{J}.}
\label{f5.1}
\end{figure}

\end{document}